\documentclass{aa}  

\usepackage{graphicx}
\usepackage{txfonts}

\usepackage{newtxtext,newtxmath}
\usepackage[T1]{fontenc}
\usepackage{graphicx}	
\usepackage{amsmath}	

\usepackage{xcolor}

\newcommand{\SII}{[S\,{\sc ii}]}

\newcommand{\HII}{H\,{\sc ii}}

\newcommand{\Ha}{H$\alpha$}
\newcommand{\Hb}{H$\beta$}

\newcommand{\Msun}{~M$_{\odot}$}

\begin{document}

   \title{Metal-THINGS: The association and optical characterization of SNRs with H\,{\sc i}  holes in NGC 6946}
   \titlerunning{The association of SNR with  H\,{\sc i}  holes in NGC~6946}


   \author{M. A. Lara-L\'opez \inst{1}
          \and
        L.~S.~Pilyugin\inst{2,3}
          \and
        J.~Zaragoza-Cardiel\inst{4,5}
        \and
        I. A. Zinchenko\inst{6,3}
                \and{ O.~L\'{o}pez-Cruz}\inst{4}
                \and{S. P. O'Sullivan}\inst{7}
\and{M.~E.~De Rossi}\inst{8,9}
       \and{S.~Dib}\inst{10}
        \and{L. E. Gardu\~{n}o}\inst{4}
        \and{M.~Rosado}\inst{11}
        \and{ M. S\'anchez-Cruces}\inst{11}
        \and{ M. Valerdi}\inst{4}
                  }

   \institute{Departamento de F\'{ı}sica de la Tierra y Astrof\'{ı}sica, Instituto de F\'{ı}sica de Part\'{ı}culas y del Cosmos, IPARCOS. Universidad Complutense   de Madrid (UCM), E-28040, Madrid, Spain
              \email{maritzal@ucm.es}
         \and
             Institute of Theoretical Physics and Astronomy, Vilnius University, Sauletekio av. 3, 10257, Vilnius, Lithuania
             \and
              Main Astronomical Observatory, National Academy of Sciences of Ukraine, 27 Akademika Zabolotnoho St, 03680, Kiev, Ukraine
              \and
Instituto Nacional de Astrof\'{ı}sica,  \'{O}ptica y Electr\'{o}nica (INAOE), Luis E. Erro 1, Tonantzintla, Puebla, C.P. 72840, M\'{e}xico
                    \and                
Consejo Nacional de Ciencia y Tecnolog\'{i}a, Av. Insurgentes Sur 1582, 03940, Ciudad de M\'{e}xico, Mexico
              \and
Faculty of Physics, Ludwig-Maximilians-Universit{\"a}t, Scheinerstr. 1, 81679 Munich, Germany
              \and
School of Physical Sciences and Centre for Astrophysics \& Relativity, Dublin City University, Glasnevin, D09 W6Y4, Ireland
              \and
Universidad de Buenos Aires, Facultad de Ciencias Exactas y Naturales y Ciclo B\'{a}sico Com\'{u}n. Buenos Aires, Argentina
              \and
CONICET-Universidad de Buenos Aires, Instituto de Astronom\'{i}a y F\'{i}sica del Espacio (IAFE). Buenos Aires, Argentina
             \and
Max Planck Institute for Astronomy, K\"{o}nigstuhl 17, 69117, Heidelberg, Germany
              \and
Instituto de Astronom\'{i}a, Universidad Nacional Autonoma de M\'{e}xico, Apartado Postal 70-264, CP 04510 M\'{e}xico, CDMX, M\'{e}xico
 }

   \date{Received TBD; accepted TBD}

 
  \abstract
   { NGC~6946, also known as the `Fireworks' galaxy, is an unusual galaxy that hosts a total of 225 supernova remnant (SNR) candidates, including 147 optically identified with high  [S\,{\sc ii}]/H$\alpha$ line ratios. In addition, this galaxy shows prominent H\,{\sc i} holes, which were analyzed in previous studies.  Indeed, the connection between SNRs and H\,{\sc i} holes together with their physical implications in the surrounding gas is worth of attention.}
   { This paper explores the connection between the SNRs and the H\,{\sc i} holes, including an analysis of their physical link to observational optical properties inside and around the rims of the holes, using new integral field unit (IFU) data from the Metal-THINGS survey.}
   {We present an analysis combining previously identified  H\,{\sc i} holes, SNRs candidates, and new integral field unit (IFU) data from Metal-THINGS of the spiral galaxy NGC 6946, which has an unusually high supernova rate.  
  We analyze the distributions of the oxygen abundance, star formation rate surface density, extinction, ionization,  diffuse ionized gas, and the Baldwin-Phillips-Terlevich classification throughout the galaxy. }
   { By analyzing in detail the optical properties of the 121 previously identify H\,{\sc i} holes in NGC 6946,  we find that the SNRs are concentrated at the rims of the H\,{\sc i} holes. Furthermore, our IFU data shows that the star formation rate and extinction are enhanced at the  rims of the holes. To a lesser degree, the oxygen abundance and ionization parameter show hints of enhancement on the rims of the holes.
 Altogether, this provides evidence of induced star formation taking place at the  rims of the holes, whose origin can be explained by the expansion of superbubbles created by multiple supernova explosions in large stellar clusters dozens of Myr ago.}{}


   \maketitle
%

\section{Introduction}

NGC~6946, also known as the "Fireworks galaxy" due to its high number of supernovae (SNe), is a nearly face-on galaxy, with a reported inclination angle between $\sim$38$\degr$ \citep{Carignan1990,Boomsma2008}, and  $\sim$32$\degr$ \citep{Bonnarel1988,deBlok2008}.

Morphologically, NGC~6946 is a late-type SABcd galaxy \citep{RC3} and hosts well-defined multiple spiral arms and three bars \citep[e.g.,][]{Elmegreen1998,Schinnerer2006,Fathi2009,Romeo2015}. Deep H\,{\sc i} observations reveal the existence of regular and prominent spiral arms in the outer gaseous disc, well outside the optical disc \citep{Boomsma2008,Bertin2010}.  In addition, NGC~6946  is an isolated galaxy located in the nearby void \citep{Sharina1997}, although the presence of tidal encounters and minor mergers cannot be discarded \citep{Boomsma2008}

Due to its intriguing nature, NGC~6946 has been  mapped at multiple wavelengths. For example, in the 21 cm atomic hydrogen emission line \citep[e.g., ][]{Rogstad1972,Carignan1990,Boomsma2007,deBlok2008}, in  H$\alpha$ emission
 of the ionized hydrogen \citep[e.g., ][]{Bonnarel1988,BlaisOuellette2004,Daigle2006},  in the molecular (CO) emission lines \citep[e.g., ][]{Walsh2002, Leroy09,Helfer03,Romeo2015}, with the Hubble Space Telescope (HST) \citep{Larsen04}, and as part of large surveys, such as the Key Insights on Nearby Galaxies: a Far-Infrared Survey with Herschel  \citep[KINGFISH,][]{Kennicutt11}, and 
the Spitzer Infrared Nearby Galaxies Survey \citep[SINGS,][]{Moustakas10}, among others.

In particular,  NGC~6946 is part of  the  H\,{\sc i} Nearby Galaxy Survey (THINGS)  survey \citep{Walter2008}, which observed 34 large,
nearby galaxies with the Very Large Array (VLA), obtaining high spatial and spectral resolution  H\,{\sc i} data. Therefore, the distribution of
properties (surface mass density, velocity dispersion, and rotation velocity) of the  atomic gas across the disc are available for this galaxy.

NGC~6946 is the most extreme known example of a galaxy with a high supernova rate. Ten supernova events have been observed since 1917 \citep{Eldridge2019}.  A large number of supernova remnants (SNR) and candidates have been identified in  NGC~6946 in different ways \citep{Matonick1997,Long2019,Long2020}.  
Recently, a large optical search of SNR was performed by  \citet{Long2019} using the WIYN telescope (Kitt Peak National Observatory). 
A total of 225 SNR candidates were found by \citet{Long2020} using optical and IR data, including 147 SNRs identified as optical nebulae with high [S\,{\sc ii}/H$\alpha$] line ratios  \citep{Long2019}.

One  prominent feature detectable with H\,{\sc i} observations is the existence of  H\,{\sc i} holes \citep{Brinks86}. \citet{Boomsma2007} and \citet{Boomsma2008} studied the distribution and kinematics of the neutral gas in NGC 6946 at 21-cm
with the Westerbork Synthesis Radio Telescope. They detected high-velocity H\,{\sc i} gas and found 121  H\,{\sc i} holes, most of which
are located in the inner regions where the gas density and the star formation rate are higher. 
The sizes  of the H\,{\sc i} holes are up to 3 kpc diameter, while their ages are in
the range of 1 to 6$\times$10$^{7}$ yr.\footnote{\citet{Boomsma2007} and \citet{Boomsma2008} adopt a distance of 6 Mpc to  NGC~6946, while recent distance estimates using the tip of the red giant branch result in a value of around 7.8 Mpc \citep{Anand2018,Murphy2018}. This implies that the sizes and ages of  the H\,{\sc i} holes can be underestimated  by around thirty per cent.}   
\citet{Boomsma2007} and \citet{Boomsma2008} conclude that stellar feedback (galactic fountain) is probably at the origin of most of the high-velocity gas
and   H\,{\sc i} holes \citep[see also][]{Dib21}. That is, that the most likely mechanism for producing H\,{\sc i}  holes is the expansion (and blow-up) of superbubbles created by
multiple supernova explosions around large stellar clusters. 

The number of holes in NGC~6946 and their sizes  is still debatable. A later study by \citet{Bagetakos11} found only 56 holes, although with an excellent agreement for the holes larger than 700 pc in comparison with \citet{Boomsma2008}. Part of the disagreement can be attributed to the different resolution limit of each study.  Moreover, both studies rely on a "by eye" identification of the  H\,{\sc i}  holes,  which is  less quantifiable  and  prone to human error.


We observed NGC 6946 as part of the Metal-THINGS survey with integral field unit (IFU) spectroscopy.  A mosaic was built with a collection of 12 pointings, which is, to the best of our knowledge, the largest collection of optical spectra of this galaxy to date. Our aim is to characterize the optical properties of this galaxy such as the oxygen abundance gradient and  the SFR, as well as to characterise the SNRs and their possible effect on the aforementioned properties. Due to the starburst nature of this galaxy, we search as well for relations between the observed H\,{\sc i} holes, SNRs, and star formation rate (SFR) enhancements.
 Throughout this paper, we use the most recent distance estimate using the tip of the red giant branch of 7.8 Mpc \citep{Anand2018,Murphy2018}.




This paper is organized as follows. The observations  are described in Section 2. In Section 3, we determine the
oxygen abundance, SFR surface density, ionization parameter and color excess in  NGC~6946. In Section 4, our results are presented, and Section 5 provides a Summary and Conclusions.


\section{Observations}\label{sec:obs}

The Metal-THINGS survey is a large program focussed on obtaining IFU spectroscopy of
a unique sample of nearby galaxies with complementary imaging and spectroscopic mapping at multiple
wavelengths. Our sample of galaxies is based on the THINGS survey \citep{Walter2008}, which observed 34 large,
nearby radio galaxies with the Very Large Array (VLA), providing high spatial and spectral resolution H\,{\sc i}  data.

The Metal-THINGS survey is obtaining IFU data using the 2.7m Harlan Schmidt telescope at McDonald Observatory,
with the George Mitchel spectrograph \citep[GMS, formerly known as VIRUS-P,][]{Hill2008}.
 GMS is a square array of 100 $\times$ 102 arcsec. The IFU consists of 246 fibers arranged in a fixed pattern, where each fiber 
has a 4.2 arcsec diameter.
Every pointing is observed with 3 dither positions to ensure a 90$\%$ surface coverage \citep[see Fig. 1 of ][]{Blanc09}. Due to the extended nature
of our galaxies, sky exposures are taken off-source for sky subtraction purposes during the reduction process.

NGC~6946 was observed as part of the Metal-THINGS survey
 \citep{LaraLopez2021} through a red setup covering the wavelength range from 4400 to 6800 \AA,  with the low resolution grid (VP1), which provides a spectral resolution of  $\sim$5 {\AA} FWHM.
The wavelength range of the red setup allows us to measure the strongest emission lines:  H$\beta$, [OIII]$\lambda\lambda$4959, 5007,
[OI]$\lambda$6300, H$\alpha$, [NII]$\lambda$6584, [SII]$\lambda\lambda$6716, 6731. 

Our observing procedure consists on 15 mins exposure per dither, followed by a sky exposure, and repetition of
the process until 45 mins are reached per dither, per pointing. A calibration star was observed every night using 6
dither positions to ensure a 99$\%$ flux coverage. Calibration lamps (Neon $+$ Argon for the red setup) were observed
at the beginning and end of every night to wavelength calibrate the spectra.

\begin{figure}
\resizebox{1.00\hsize}{!}{\includegraphics[angle=000]{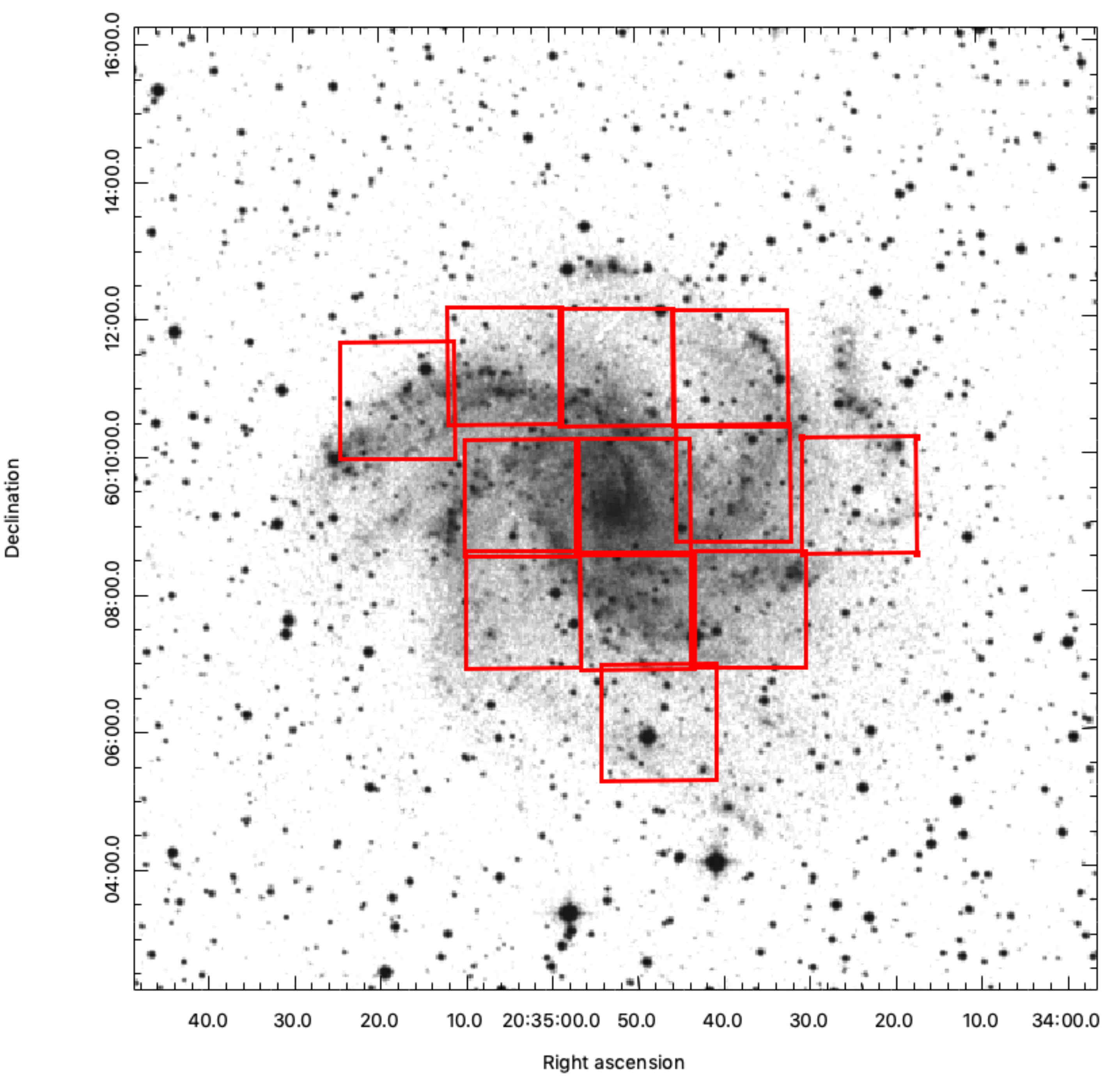}}
\caption{
Observed 12 pointings superimposed on the NGC~6946 image. Each individual box corresponds to the field of view of the George Mitchel spectrograph (GMS, formerly known as VIRUS-P).
} 
\label{figure:pointings}
\end{figure}

NGC~6946 was observed using GMS in June 2018, October 2018, October 2020, and June 2021. The average seeing during the observations was 1.5 arcsecs. A total of 12 pointings were observed to form the mosaic shown in Fig \ref{figure:pointings}. The whole mosaic has a total of 8,856 individual spectra. 
The astrometry was set by using all the fibers of individual pointings and their relative positions within each. First, we created a tridimensional cube associating each fiber to its relative spatial position in the pixel (x,y) plane and convolving by a gaussian with a FWHM of 4$\farcs$2 (the diameter of the fiber) at each wavelength. Next, from the spectral-dimension collapsed image we identified several stars and used the same star positions from 2MASS images to set an absolute astrometry for each pointing. We repeated this process for each pointing to create the mosaic of Fig \ref{figure:pointings}. Even though our initial pointings were aligned next to each other, since the observing runs spread over several years, we attribute the observed shifts to variations in the instrumentation setups. 


The galaxy NGC 6946 is at a distance of about 7.8 Mpc \citep{Anand2018,Murphy2018} which gives a scale of 37.8 pc/$\arcsec$,
so that each 4$\farcs$2 fiber in the IFU corresponds to $\sim$158.8 pc.
The optical radius of NGC~6946 ($R_{25}$ = 344.45 arcsec) is taken  from the NASA/IPAC Extragalactic Database
({\sc ned})\footnote{The NASA/IPAC Extragalactic Database ({\sc ned}) is
operated by the Jet Populsion Laboratory, California Institute of Technology, under contract with the National Aeronautics and Space
Administration.  {\tt http://ned.ipac.caltech.edu/} }. 
With the adopted distance, this gives an optical radius of $R_{25}$ = 13.02 kpc. 

The basic data reduction including bias subtraction, flat frame correction and wavelength calibration, was performed using
P3D \footnote{https://p3d.sourceforge.io}. The sky subtraction and the combination of dithers was performed using our own routines
in Python. Flux calibration was performed following \citet{Cairos2012}  using 6 dither positions, ensuring 99$\%$ coverage.
Next, we used IRAF \citep{Tody1986} to create a flux calibration function using the packages \textit{standard} and \textit{sensfunc}.

The stellar continuum of all flux-calibrated spectra was fitted using STARLIGHT \citep{CidFernandes2005, Mateus2006, Asari2007}.
Briefly,  we used 45  simple stellar population (SSP) models from the evolutionary synthesis models of
\citet{Bruzual2003} with ages from 1 Myr up to 13 Gyr and metallicities Z=0.005, 0.02 and 0.05. 
The reddening law of \citet{Cardelli1989} was adopted, assuming R$_{\rm v}$=3.1.
Next, the fitted  continuum  was subtracted from the spectra,
and the emission lines were measured using Gaussian line-profile fittings. It should be noted that only those fiber spectra where
all the used lines were measured with a signal-to-noise ratio S/N $> 3$ were considered. For a more detailed description, see \citet{Zinchenko2016}.

In this paper we are using the individual spectra of the fibers, which are independent of each other. This is in contrast to the spaxel spectra in surveys such as the Mapping Nearby Galaxies at
Apache Point Observatory survey (MaNGA), where the point spread function (PSF) is estimated to have a FWHM of
2.5$\arcsec$  \citep{Bundy2015,Belfiore2017}. Our IFU spectroscopy and derived astrometry provide the possibility to construct the maps of the
oxygen abundance (and other characteristics) in the disc and to investigate the global and local properties of the abundance
distribution \citep{LaraLopez2021}.


\section{Physical properties of  NGC~6946}



The geometrical parameters of the galaxy (location of the center of the galaxy, the position angle of the major axis and the inclination angle)  
are needed to determine  the galactocentric distances of the fibers. 
In this paper, we use the position of the center  obtained by \citet{Trachternach2008}  (RA = 20$^{h}$ 34$^{m}$ 52$\fs$2 and
DEC = +60$\degr$ 09$\arcmin$ 14$\farcs$4),  and the position angle of the major kinematic axis (PA = 242$\fdg$7) and the inclination angle ($i$ = 32$\fdg$6)  obtained by \citet{deBlok2008}.

 The stellar mass estimates for this galaxy show an appreciable spread. At the  distance
of 5.9 Mpc, using data from the {\it Spitzer} Infrared Nearby Galaxies Survey,
 \citet{deBlok2008} found the mass value to be within the range log(M$_{\star}$/M$_{\sun}$) $\sim$ 10.58 -- 10.77. 
On the other hand, \citet{Jarrett2019} found a value of  log(M$_{\star}$/M$_{\sun}$)= 10.26 using the mid-infrared luminosity from the Wide-field Infrared Survey Explorer ({\em WISE}). NGC~6946 does not show a classical bulge but has a tiny pseudobulge, and there is no evidence that this galaxy hosts an AGN \citep{Kormendy2010}.

\subsection[]{Diffuse ionized gas}\label{DIG}

\begin{figure}
\resizebox{1.0\hsize}{!}{\includegraphics[angle=000]{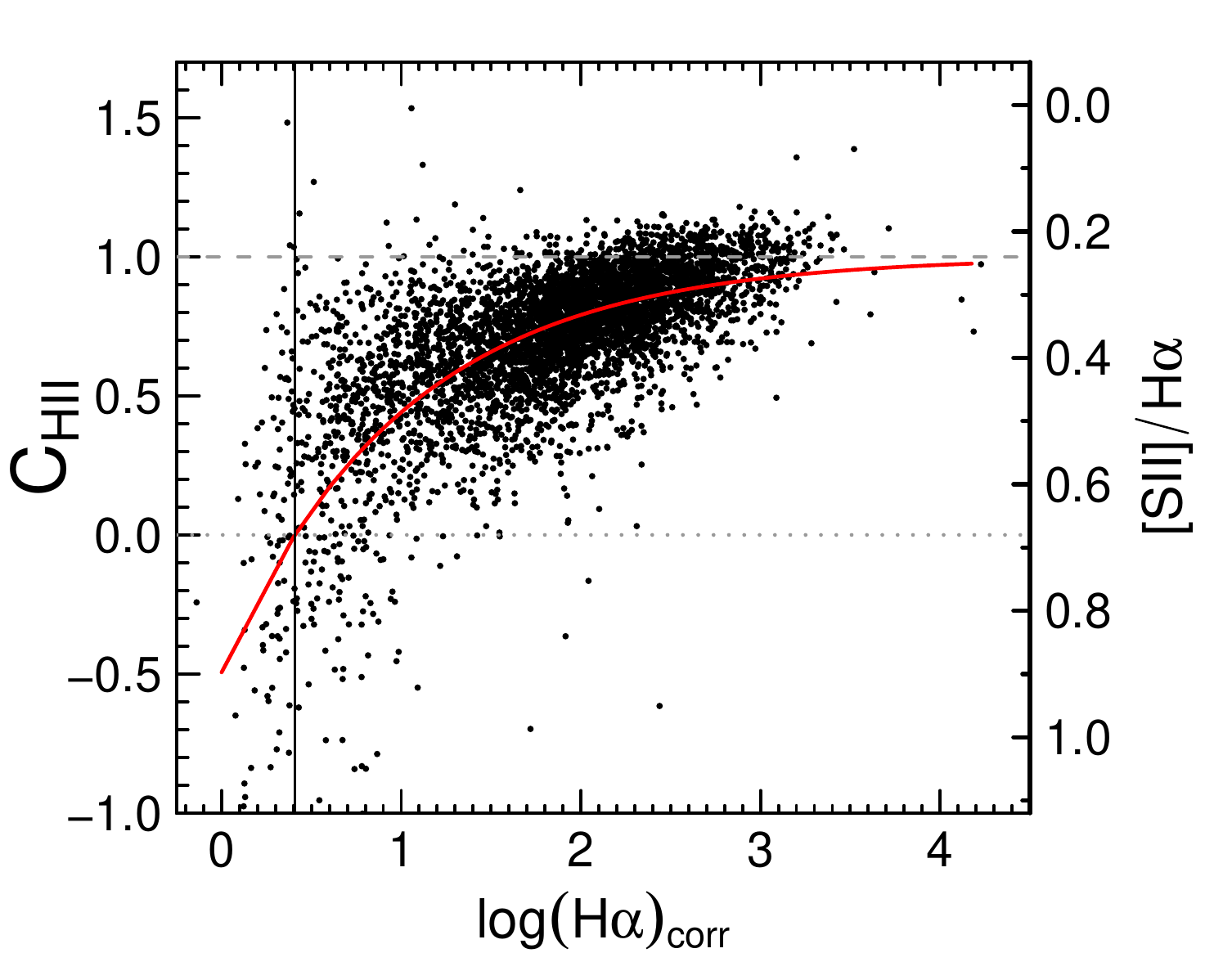}}
\caption{Identification of DIG based on the extinction corrected \Ha\ flux and ${\rm C}_{\rm H\,{\textsc{ii}}}$. The dashed and dotted horizontal lines show the median \SII /\Ha\  for the 100 brightest and 100 dimmest fibers, respectively. The vertical solid line shows the value of log($f_0$). The red solid line shows the fit of Eq. \ref{CHIIf} to our data. \label{fig:DIG_Rel}}
\end{figure}

 The diffuse ionized gas (DIG), also known as the warm ionized medium (WIM), is a warm ($\sim$10$^4$ K), low density ($\sim$10$^{-1}$cm$^{-3}$ ) gas phase found in the interstellar medium of galaxies  \citep{Reynolds84}. DIG is an important component in star-forming galaxies, and its contribution to emission line fluxes can affect and impact the interpretation of measured metallicities, ionization, and 
BPT diagrams \citep[][]{Baldwin1981} in galaxies.   Several approaches have  been proposed for the identification of DIG using IFU spectroscopy \citep[e.g.][]{Poetrodjojo19,Kaplan16}. \\ 

To determine the fraction of flux originating from H\,{\textsc{ii}}  regions and from the DIG (named C$_{\rm H\,{\textsc{ii}}}$), we use the method first proposed by \citet{Blanc09}, and further developed by \citet{Kaplan16}. This method relies on the assumption that the brightest fiber fluxes are dominated by H\,{\textsc{ii}}  emission, while the dimmest ones are dominated by DIG emission. Following \citet{Kaplan16}, we first estimate the characteristic value of \SII /\Ha\  in the brightest fibers for  \HII\  regions (\SII /\Ha)$_{\rm H\,{\textsc{ii}}}$, and in the dimmest ones for DIG  (\SII /\Ha)$_{\rm \scriptscriptstyle DIG}$. Only fibers with a S/N $>$ 3 in \Ha, \Hb, and \SII\ are included in this analysis. 

Then, in each fiber, we make an   initial guess for C$_{\rm H\,{\textsc{ii}}}$:

\begin{equation}\label{CHII}
{\rm C}_{\rm H\,{\textsc{ii}}}=  \frac{[\rm S\,{\textsc{ii}}] / \rm H\alpha - ([\rm S\,{\textsc{ii}}] / \rm H\alpha)_{\rm \scriptscriptstyle DIG}} {([\rm S\,{\textsc{ii}}] / \rm H\alpha)_{\rm \scriptsize H\,{\textsc{ii}}} - ([\rm S\,{\textsc{ii}}] / \rm H\alpha)_{\rm \scriptscriptstyle DIG} }
\end{equation} 

Next, by using the initial estimate of ${\rm C}_{\rm H\,{\textsc{ii}}}$ and the extinction corrected $f({\rm H{\alpha}})$, we fit the data for $f_0$ and $\beta$ in the following expression:

\begin{equation}\label{CHIIf}
{\rm C}_{\rm H\,{\textsc{ii}}}=  1- \left({\frac{{ f_0}}{f{\rm (H{\alpha})}} } \right)^{\beta}
\end{equation} 

obtaining $f_0$ =    2.56  $\pm$  0.12, and $\beta$= 0.427  $\pm$   0.007, shown by the red solid line in Fig. \ref{fig:DIG_Rel}. A final value for ${\rm C}_{\rm H\,{\textsc{ii}}}$ is provided by Eq. \ref{CHIIf} using the best-fit coefficients. 

While all regions where \Ha$_{\rm corr}$ $<$ $f_0$ are 100$\%$ DIG dominated, a somewhat arbitrary threshold  in ${\rm C}_{\rm H\,{\textsc{ii}}}$ is needed to select the DIG, which is based on the dimmest fluxes in the observed sample. In this paper, we define fibers with ${\rm C}_{\rm H\,{\textsc{ii}}}$ $<$ 0.4 as dominated by DIG, which are flagged in the rest of our analysis. In Fig. \ref{figure:bptmap}, we show the fibers identified as DIG in small black circles, which corresponds to  9.2$\%$ of the analyzed sample.

\subsection{BPT classification}

\begin{figure}
\resizebox{1.0\hsize}{!}{\includegraphics[angle=000]{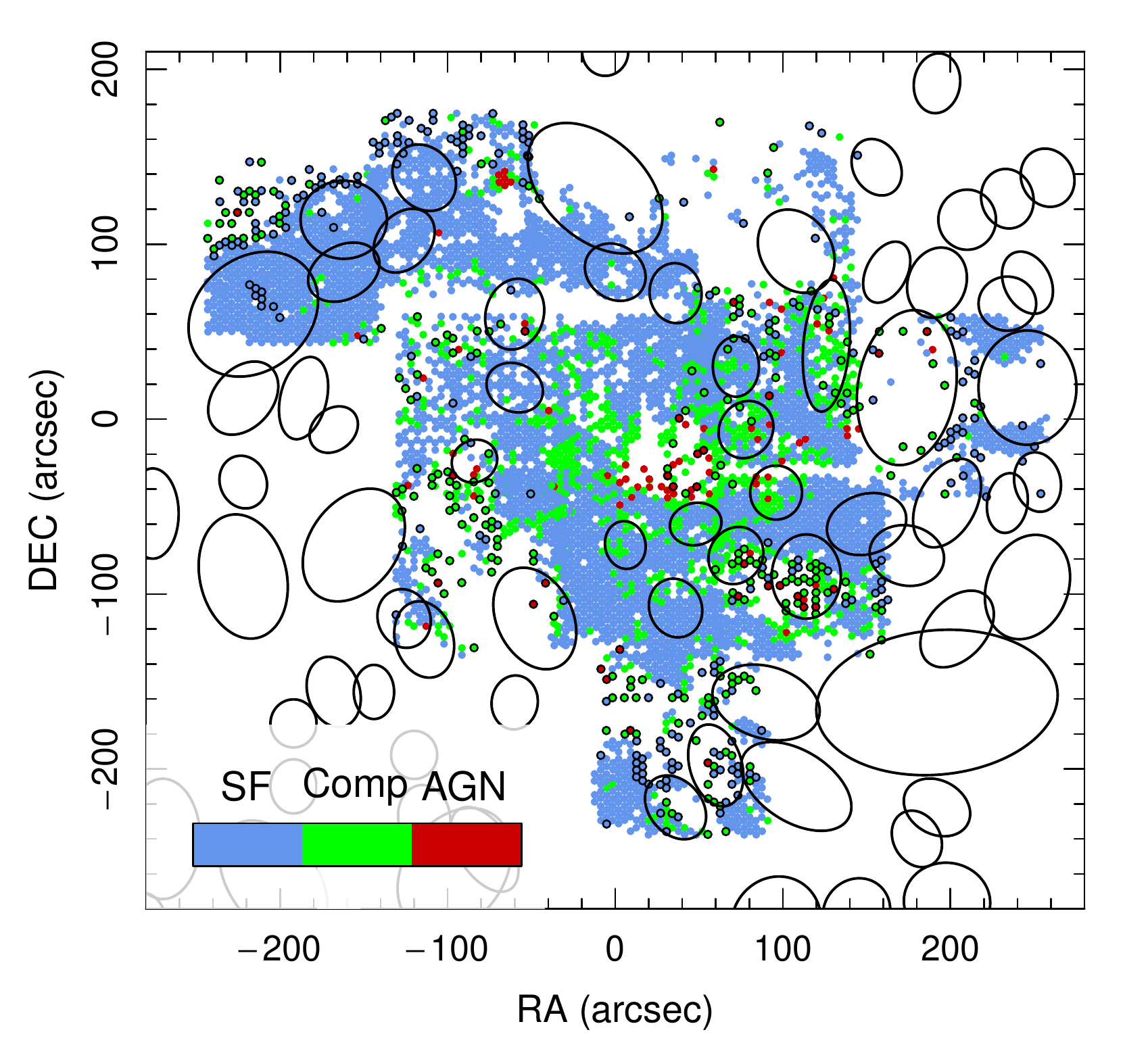}}
\resizebox{1.0\hsize}{!}{\includegraphics[angle=000]{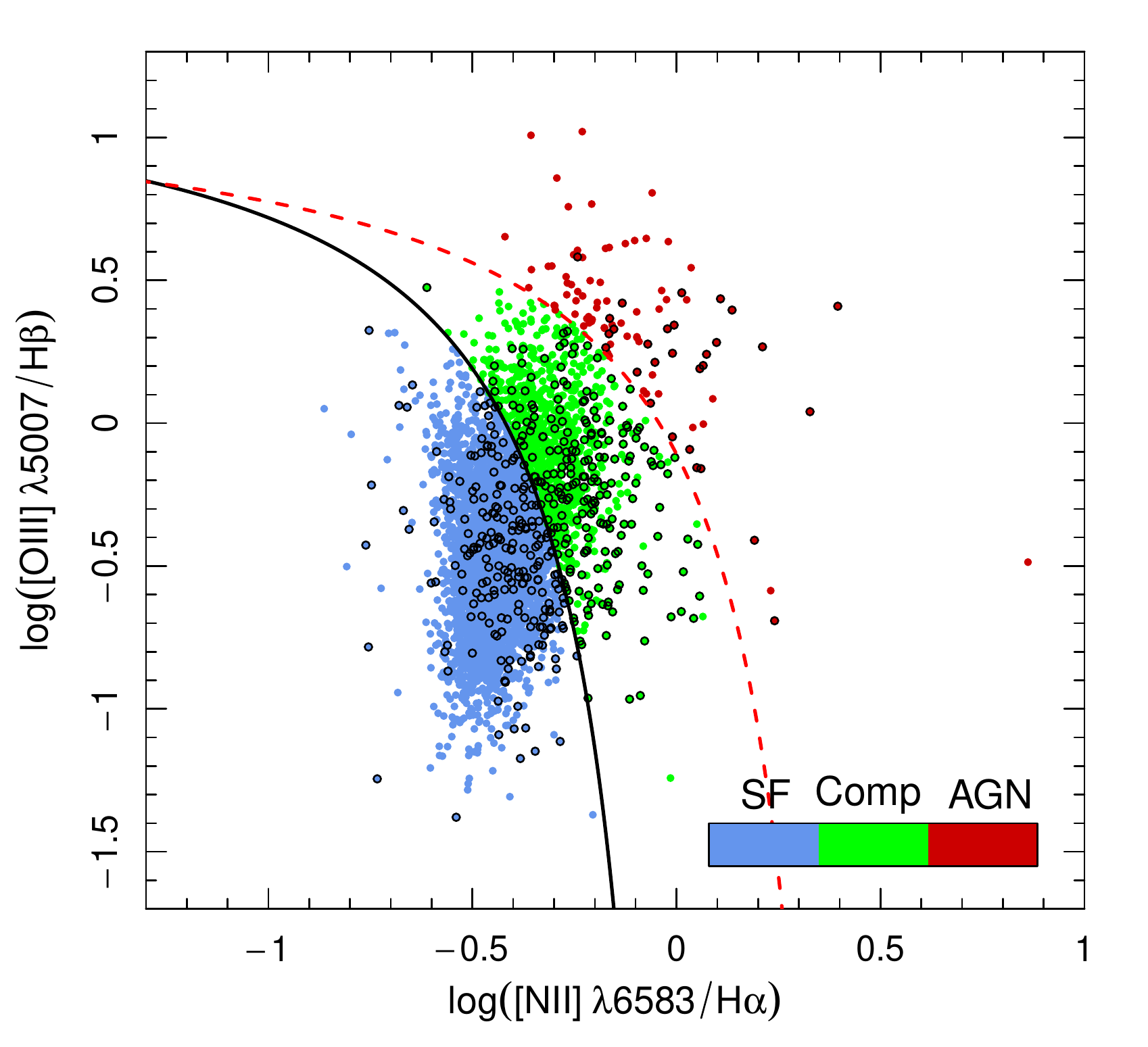}}
\caption{{\em Top panel}: map of the BPT classification, color-coded as indicated in the label.
{\em Bottom Panel}:  BPT diagram for the individual fibers with SF-like spectra (blue symbols), Composite (green symbols), and AGN-like (red
symbols). The black solid and red dashed curves are the demarcation lines between SF and AGN regions defined by \citet{Kauffmann2003}
and \citet{Kewley2001}, respectively. The ellipses in the top panel indicate the H\,{\sc i}  holes (see \S \ref{SecSNRyHoles}), while the small black circles denote regions dominated by DIG.} 
\label{figure:bptmap}
\end{figure}


The measured emission line fluxes of each fiber spectra are corrected for the interstellar reddening using the reddening law from \citet{Cardelli1989} with $R_{V}$ = 3.1.
The value of the interstellar reddening is estimated through the comparison between the measured and the theoretical H$\alpha$/H$\beta$ ratio
(that is, using the standard  value of H$\alpha$/H$\beta$ = 2.86). When the measured value of H$\alpha$/H$\beta$ is lower than 2.86, the reddening is assumed to be zero. 
We adopted $C_{{\rm H}{\beta}} = 0.47A_{V}$ \citep{Lee2005}.

 We classify every spectra using its location in the [N\,{\sc ii}]$\lambda$6584/H$\alpha$ versus [O\,{\sc iii}]$\lambda$5007/H$\beta$
line ratio diagram suggested by \citet[][BPT diagram]{Baldwin1981},  The NGC~6946  map and BPT classification  is shown in Fig.~\ref{figure:bptmap}.
As in our previous studies \citep{Zinchenko2019,Pilyugin2020a,Pilyugin2020b,Pilyugin2021a,Pilyugin2021b,LaraLopez2013},
the spectra located to the left (below) the demarcation line from \citet{Kauffmann2003}  (blue symbols in the two panels  of Fig.~\ref{figure:bptmap}) are referred to as
the SF-like or H\,{\sc ii} region-like spectra. 
Those located to the right (above) the demarcation line from \citet{Kewley2001} (red symbols in Fig.~\ref{figure:bptmap}) are referred to as the AGN-like spectra. The spectra located between those demarcation lines are  Composite, or intermediate spectra
(green symbols in Fig.~\ref{figure:bptmap}).  In both panels of Fig.~\ref{figure:bptmap}, the black circles denote regions dominated by DIG.

\subsection{Oxygen abundance and Ionization parameter}\label{OxygenAbundanceMeasure}

Here we estimate the oxygen abundance for the fibers classified as H\,{\sc ii} region-like by the BPT. The oxygen lines [O\,{\sc ii}]$\lambda$3726 and $\lambda$3729 are beyond the spectral range of our red GMS spectra. Hence, the calibrations which involve those lines cannot be used for abundance determinations with our current data.
From the plethora of available methods, we use the three-dimensional S calibration O/H = $f$($N2,S2,R3/S2$) from \citet{Pilyugin2016}. The advantage of 3D calibrations for the abundance determinations in comparison to 1D calibrations is discussed in \citet{Pilyugin2018}. 
The S calibration consists of an upper (high metallicity) and lower (low metallicity) branches. 
The transition from the upper to lower branch occurs at log\,$N_{2}$ $\approx -0.6$. The corresponding calibration relation for the upper branch (log$N_{2}$ $>$ --0.6)  is  
\begin{eqnarray}
       \begin{array}{lll}
     {\rm (O/H)}^{*}_{S,U}  & = &   8.424 + 0.030 \, \log (R_{3}/S_{2}) + 0.751 \, \log N_{2}   \\  
                          & + &  (-0.349 + 0.182 \, \log (R_{3}/S_{2}) + 0.508 \log N_{2})   \\ 
                          & \times & \log S_{2},   \\ 
     \end{array}
\label{equation:ohsu}
\end{eqnarray}
where we use the notation (O/H)$^{*}_{S,U}$ = 12 +log(O/H)$_{S,U}$ for a compact way. The calibration for the lower branch
(log$N_{2}$ $<$ --0.6) is
\begin{eqnarray}
       \begin{array}{lll}
     {\rm (O/H)}^{*}_{S,L}  & = &   8.072 + 0.789 \, \log (R_{3}/S_{2}) + 0.726 \, \log N_{2}   \\  
                          & + &  (1.069 - 0.170 \, \log (R_{3}/S_{2}) + 0.022 \log N_{2})    \\ 
                          & \times & \log S_{2}   \\ 
     \end{array}
\label{equation:ohsl}
\end{eqnarray}
where (O/H)$^{*}_{S,L}$ = 12$+$log\,(O/H)$_{S,L}$. 

Since the [O\,{\sc iii}]$\lambda$5007 and $\lambda$4959 lines originate from transitions from the same energy level,  their flux ratio is defined
only by the transition probability ratio, which is close to 3 \citep{Storey2000}. The line measurement of the stronger line [O\,{\sc iii}]$\lambda$5007  
 is usually more precise than that for the weaker line [O\,{\sc iii}]$\lambda$4959. Therefore, we estimated the value of $R_3$ as
$R_3  = 1.33$~[O\,{\sc iii}]$\lambda$5007/H$\beta$. Similarly, we estimated the value of $N_2$ as $N_2 = 1.33$~[N\,{\sc ii}]$\lambda$6584/H$\beta$. The metallicity gradient and map, will be further analyzed in $\S\,\ref{SecSuperNova}$ and $\S\,\ref{SecHolesMaps}$, respectively.

 The ionization parameter is defined as  $ U = Q_{\rm ion} / 4\pi R^2_{\rm in} nc$, where $Q_{\rm ion}$ is the number of hydrogen ionizing photons emitted per second by the ionizing source, $R_{\rm in}$ is the distance from the ionization source to the inner surface of the ionized gas cloud, $n$ is the particle density, and $c$ is the speed of light.  We estimate the ionization parameter following the prescription of  \citet{Dors17}:

\begin{equation}\label{IonDors}
{ {\rm log\,(U)_D}}  = c \times S2 + d
\end{equation}

where S2 = log(\SII  $\lambda \lambda$  6717,31 / \Ha), $c = -0.26 \times (Z/Z_{\sun}) - 1.54$, and $d= -3.69 \times (Z/Z_{\sun})^2 + 5.11 \times (Z/Z_{\sun})$ -5.26. For the oxygen abundance of the Sun, we use Z$_{\sun}$=8.69 \citep{Amarsi18}. Since the oxygen abundance is needed for the ionization parameter estimation, the final sample is the same (4154 fibers).
The resulting ionization will be analyzed in   \S \ref{SecHolesMaps}.

\subsection{Star formation rates and  color excess}\label{SFRDustIonEstimation}

 In this section, we estimate the SFRs and  color excess  that will be further analyzed in detail in \S \ref{SecHolesMaps}.

The resolved SFR  is estimated using the \Ha\ prescription of \citet{Kennicutt09}  \citep[see also the review of][]{Calzetti13}, which uses an \citet{Kroupa01} initial mass function (IMF).

\begin{equation}\label{IonDors}
{ {\rm SFR [M yr^{-1}]}}  = 5.5 \times 10^{-42} L_{\rm corr} ({\rm H}\alpha_{\rm corr}) \  {\rm erg s^{-1}}
\end{equation} 

where  ${\rm H}\alpha_{\rm corr}$ is the \Ha\ flux corrected by extinction. The estimated SFRs are then converted to SFR surface densities ($\sum_{\rm SFR}$)  by dividing them by the area on the sky of the 4.2'' diameter fiber.
The SFR was measured only in fibers whose spectra are classified as SF or Composite in the BPT diagram, and have a S/N $>$ 3 in \Ha\ and \Hb. The final SFR sample used is formed by   4972 fiber spectra.

Finally, as an indicator of the dust content, we estimated the color excess E(B-V) \citep[e.g.][]{Calzetti96}. For its estimation, different ratios from Hydrogen recombination lines can be used. For our sample, we are using \Ha \ and \Hb, and the prescription of \citet{Calzetti94} in the following:

\begin{equation}
{\rm E\,(B-V)}_{\rm gas} =  {\frac{ {\rm log} ({\rm R}_{\rm obs} / {\rm R}_{\rm int}) \;  }{ 0.4 \, [ \, k (\lambda_{{\rm H}\alpha}) -  k  (\lambda_{{\rm H}\beta})\,] }}
\end{equation}

where the observed radio ${\rm R}_{\rm obs}$ is the Balmer decrement  (\Ha/\Hb), and the intrinsic ratio ${\rm R}_{\rm int}$ is 2.86, taking a case B recombination  \citep[][]{Osterbrock89}. Finally, $k(\lambda)$ corresponds to the dust extinction curve, in this case, we use \citet{Calzetti2000}, which is based on a set of starburst galaxies, appropriate for this analysis.  The color excess was estimated using the same exclusion criteria as for the SFR.

\section{Results}

\subsection{Oxygen abundance gradient and Supernova remnants}\label{SecSuperNova}

\begin{figure}
\begin{center}
\resizebox{1.0\hsize}{!}{\includegraphics[angle=000]{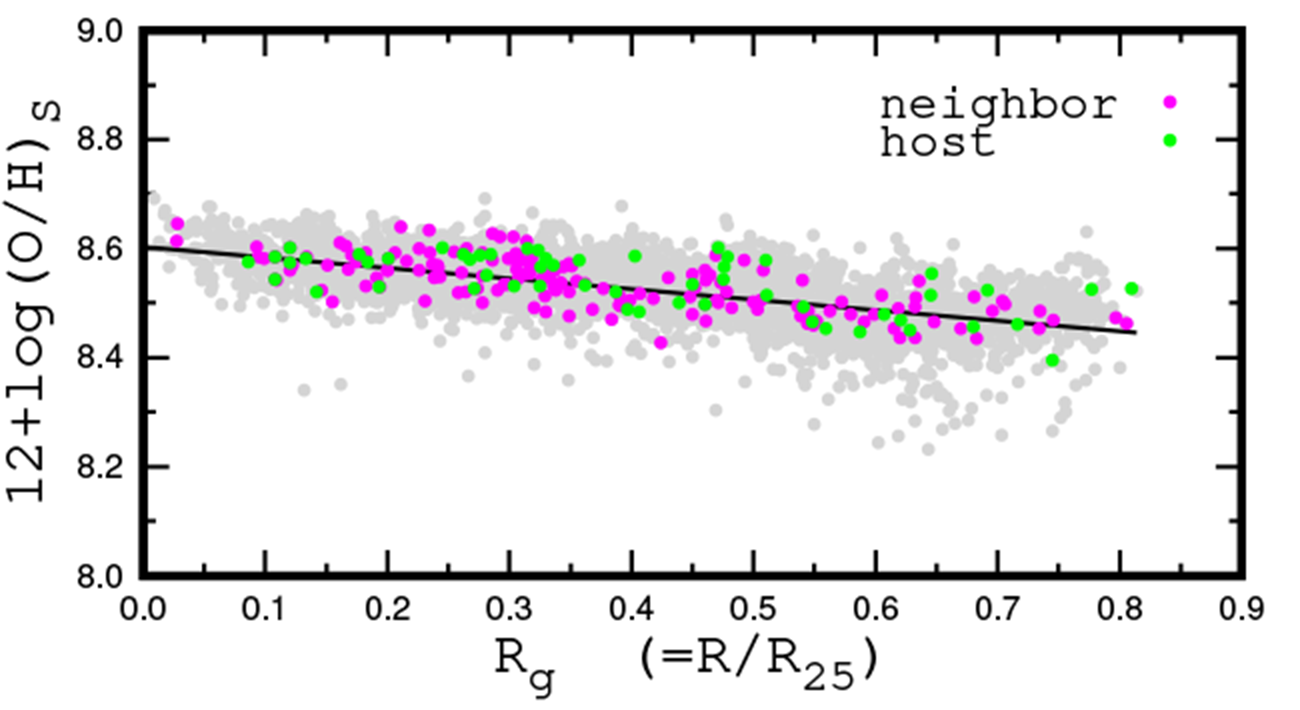}}
\caption{Radial distribution of the oxygen abundance. The grey points denote the oxygen abundance  of the individual fibers. The solid line is the best fit to the data.   The green points mark the  fibers hosting SNRs, while the magenta points correspond to the neighboring fibers to the SNRs positions.
}
\label{figure:snrohsv}
\end{center}
\end{figure}

NGC~6946 is the most extreme example of a galaxy with a high supernova rate, with ten supernova events since 1917 \citep{Eldridge2019}. 
A large number of SNR and candidates have been identified in  NGC~6946 in different ways
\citep[e.g., ][]{Matonick1997,Lacey1997,Lacey2001,Bruursema2014,Long2019,Long2020}, amounting to a total of  225 SNR candidates.  In this section, we explore whether the oxygen abundance is affected in  the immediate  environment of the SNR.

 To this aim, first we estimate the oxygen abundance gradient for the whole galaxy.  The gas metallicity estimated through strong emission lines give important clues on the physical properties of the ISM and formation and growth of galaxies. Negative gas metallicity gradients are widely found in late-type galaxies in the local universe \citep[e.g., ][]{Zaritsky94, Pilyugin2004, Moustakas10,Pilyugin2014, Belfiore2017,LaraLopez2022}. Such negative gradients are consistent with an inside-out growth scenario of the discs \citep[e.g.][]{White91, Mo98}. 

The gradient O/H = $f(R)$ is approximated by a linear relation of the type 12+log(O/H) = 12+log(O/H)$_{0}$ + grad $\times$ $R_{g}$,  where  12+log(O/H)$_{0}$ is the oxygen abundance at the center of the galaxy, grad is the gradient in units of dex/R$_{25}$, and R$_{g}$ is the radius normalised to the optical radius of the galaxy R$_{25}$. We find the following expression for the SF fibers in the disk of  NGC~6946:
\begin{equation}
12+\log{\rm (O/H)} = 8.603(\pm0.002) - 0.1938(\pm0.0037) \times R_{g} 
\label{equation:ohgradall}
\end{equation}

The mean value of the scatter around this relation is 0.0429 dex.  The solid line of Fig.~\ref{figure:snrohsv} shows the obtained relation.  In comparison with other galaxies, and considering NGC 6946 has a stellar mass of 10$^{10.26}$\Msun  \citep{Jarrett2019}, our measured gradient is in agreement with galaxies of similar stellar masses from  \citet{Kreckel19} and \citet{Zinchenko21}. A more detailed analysis on the metallicity gradients of Metal-THINGS galaxies will be presented in Lara-L\'opez et al. (in preparation).


 Next, to explore the effect of the SNRs on their enviroment, we obtain the fibers overlapping (hosting) with SNRs from the condition that the separation between the center of the fiber and the SNR
is less than the fiber radius ($r_{fiber}$ = 2.1 arcsecs), otherwise a fraction of the SNR would be beyond the field of the host fiber. 
Moreover, the uncertainty in the coordinates of the fibers is around one arcsec and could be larger for some pointings. This can result
in the incorrect identification of some fibers hosting SNRs. 
To take this possibility into account, we  also obtain a ``neighboring'' fiber(s) for each SNR from the condition that the separation between the SNR and the center of the fiber is larger than  $r_{fiber}$, and lower than $r_{fiber}$ + 2 arcsecs.
Thus, the SNR could make a contribution to the radiation in the host or neighbouring fiber, or even both, if the SNR is located in between them.

Most of host fibers are located in the SF  region of the BPT diagram. Hence, this provides  evidence that the contribution of the SNR radiation to the fiber spectra is not significant. 
Since the majority of the host and neighboring fibers show  H\,{\sc ii} like spectra,  their oxygen abundances are determined through the $S$ calibration as indicated in $\S\,\ref{OxygenAbundanceMeasure}$.   The fibers hosting SNRs are indicated as green points in Fig.~\ref{figure:snrohsv}, while the magenta points denote the neighboring fibers. It can be appreciated  that there is no systematic difference or shift between the oxygen abundance of the fibers hosting SNRs and the rest of the fibers.

Altogether, we find no appreciable difference in the properties of the gas (BPT type and abundance) in fibers hosting SNRs in comparison with the rest. This suggests that the radiation of SNRs makes a small contribution to the radiation in the fiber field.

\subsection{On the relation between SNR and H\,{\sc i} holes} \label{SecSNRyHoles}


\begin{figure}
\begin{center}
\resizebox{1.00\hsize}{!}{\includegraphics[angle=000]{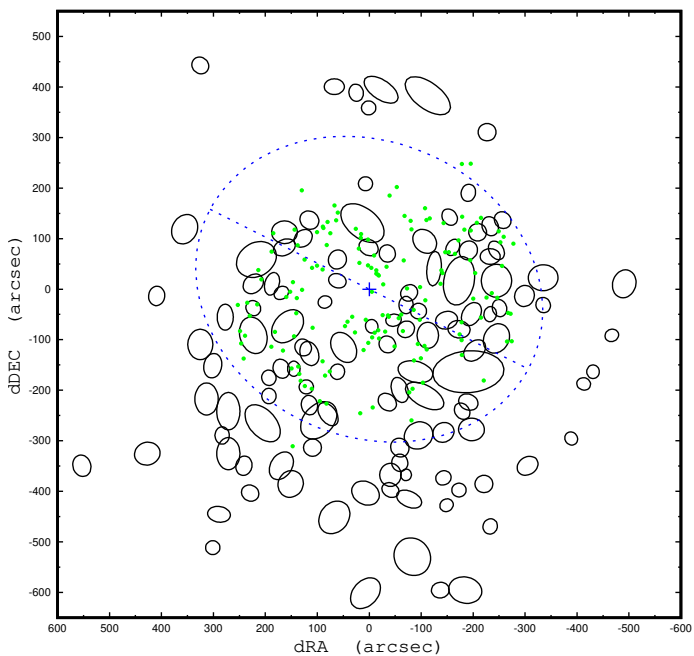}}
\caption{
  H\,{\sc i} holes and supernova remnants in the galaxy NGC~6946. The sizes and orientations of H\,{\sc i} holes are indicated by the black solid ellipses.
  The green points show the positions of the SNRs. The dashed blue ellipse denotes the optical radius of the galaxy, the plus sign marks the position of the center,
  and the dashed line denotes the position of the major axis. 
} 
\label{figure:hole-snr}
\end{center}
\end{figure}

\begin{figure}
\begin{center}
\resizebox{0.80\hsize}{!}{\includegraphics[angle=000]{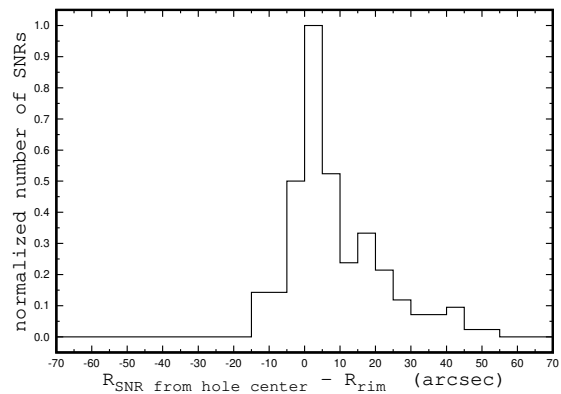}}
\caption{
  Histogram of the difference between the  SNR distance to the closest H\,{\sc i}  hole center, and the H\,{\sc i}  hole rim distance to the hole center (in the line from the center to the SNR).
} 
\label{figure:histogram-distholesSNR}
\end{center}
\end{figure}


\begin{figure*}
\begin{center}
\resizebox{0.5\hsize}{!}{\includegraphics[angle=000]{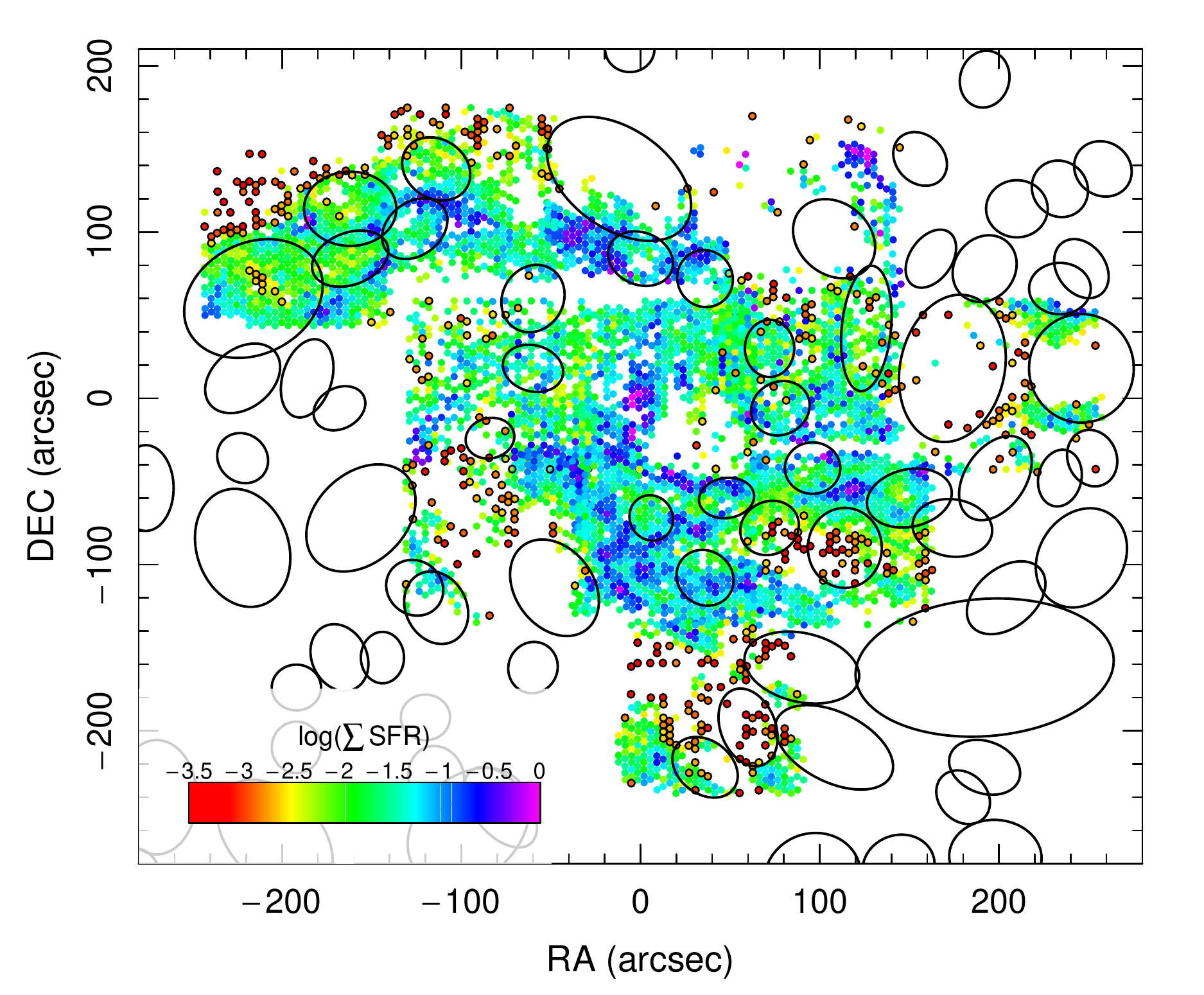}}
\resizebox{0.45\hsize}{!}{\includegraphics[angle=000]{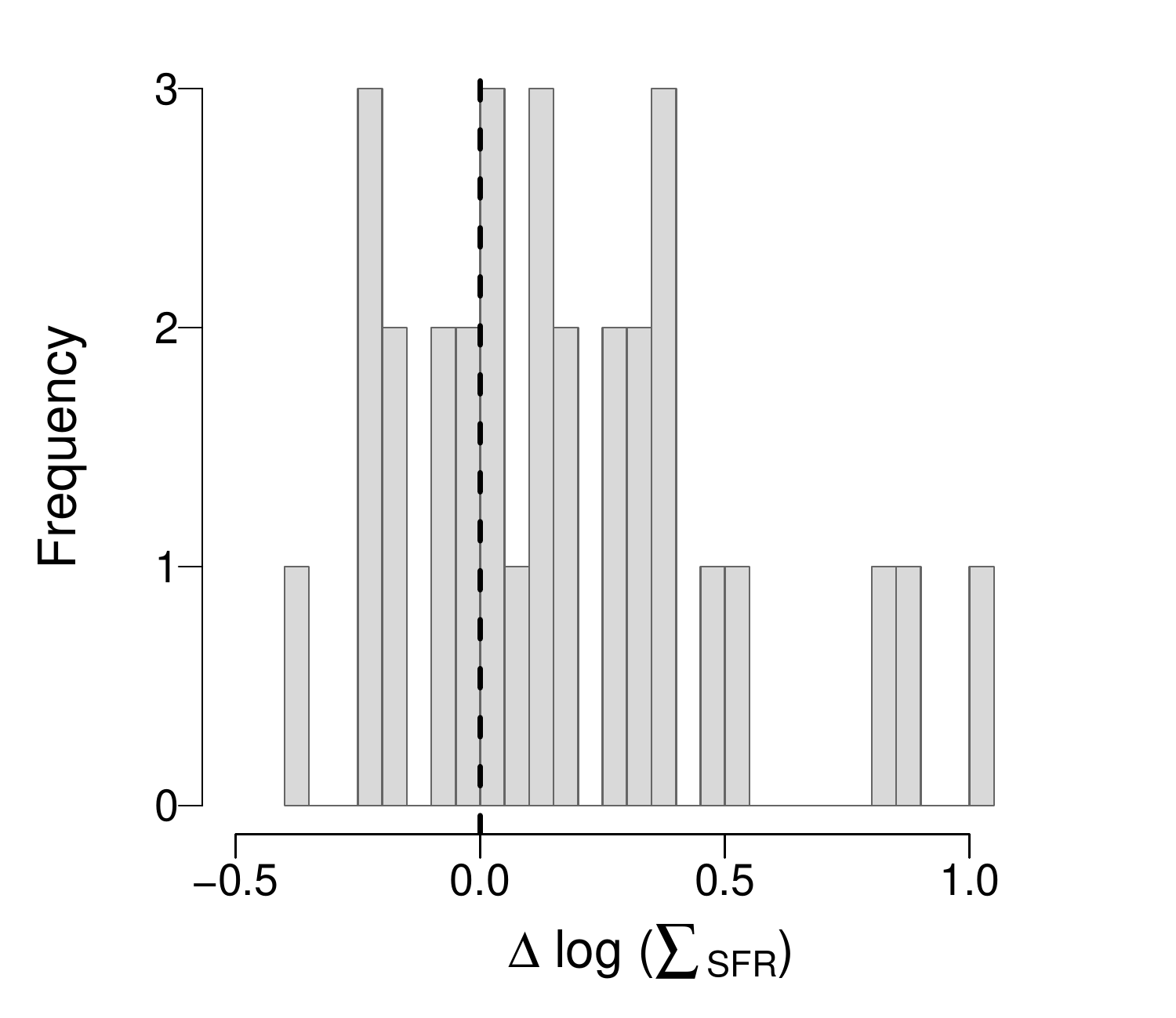}}
\caption{
  Left: H\,{\sc i} holes (black solid ellipses) superimposed on the log($\sum$$_{\rm SFR}$) map of  NGC~6946. The small black circles indicate DIG dominated fibers. Right: Histogram of the differences  $\Delta$ log($\sum$$_{\rm SFR}$) $=$ log($\sum$$\widehat{_{\rm SFR_{rim}}}$) $-$  log($\sum$$\widehat{_{\rm SFR_{hole}}}$), see text for details.} 
\label{figure:SFRMapHistHoles}
\end{center}
\end{figure*}

\begin{figure*}
\begin{center}
\resizebox{0.5\hsize}{!}{\includegraphics[angle=000]{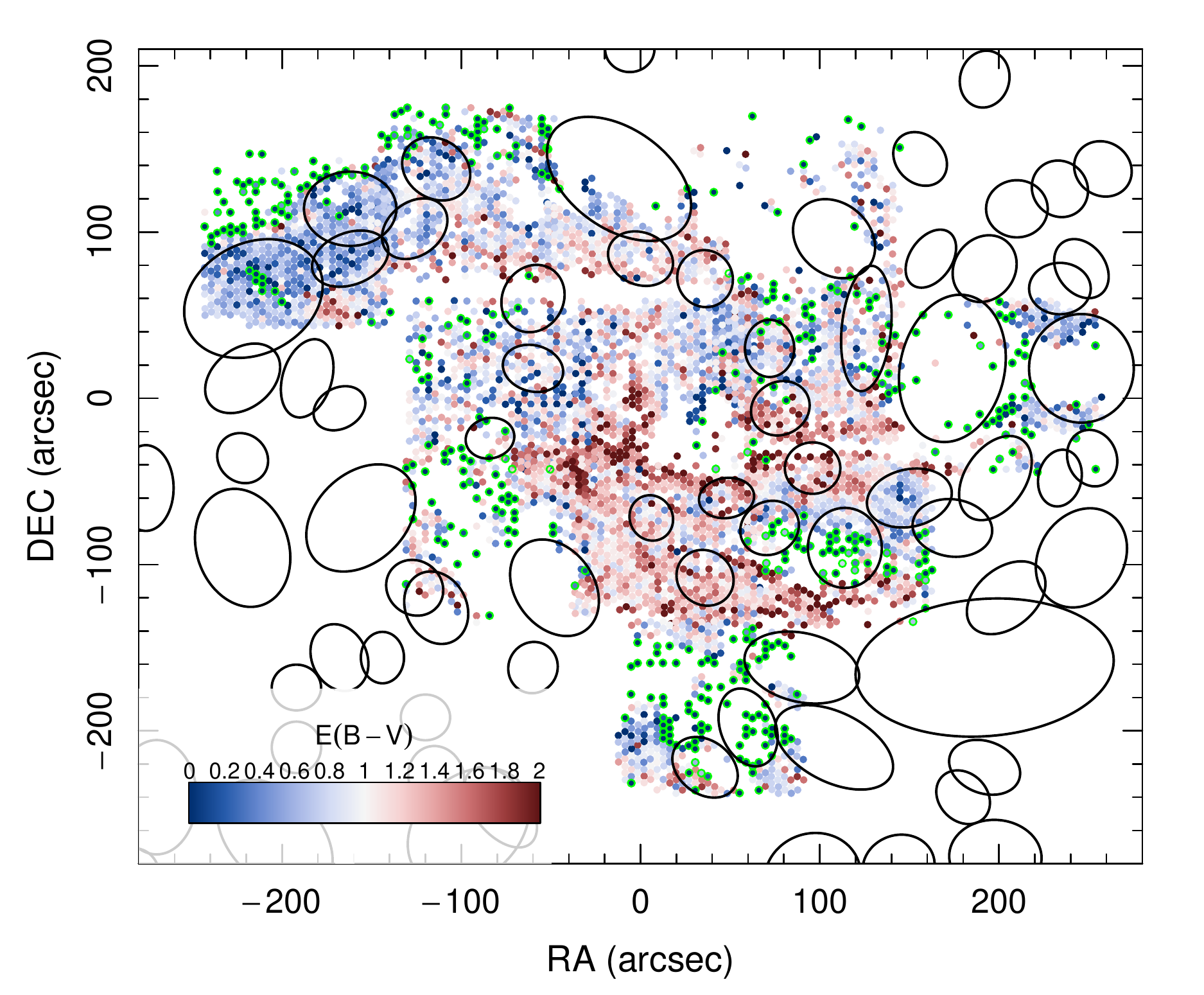}}
\resizebox{0.45\hsize}{!}{\includegraphics[angle=000]{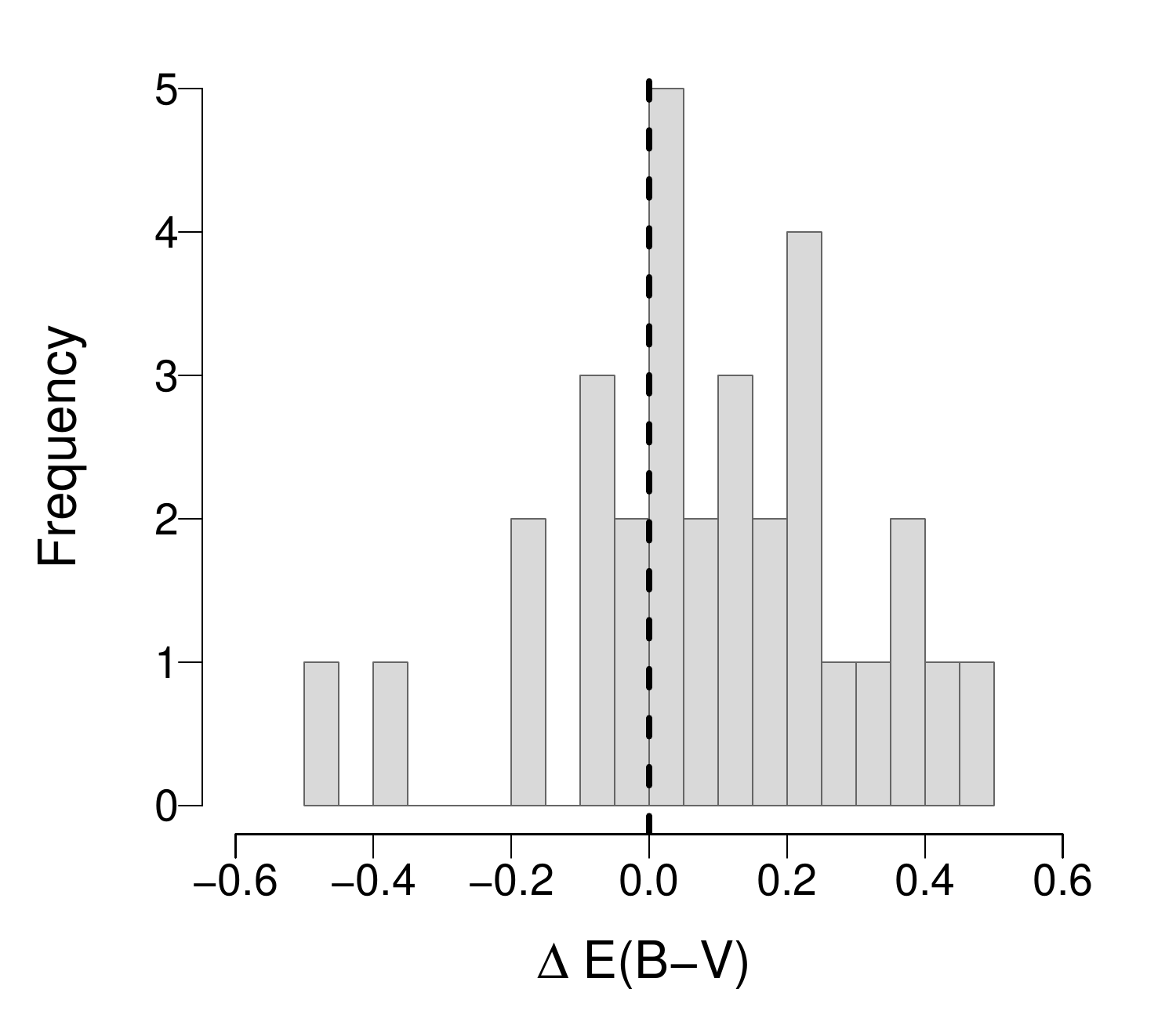}}
\caption{
  Left: H\,{\sc i} holes (black solid ellipses) superimposed on the E(B-V) map of  NGC~6946. The small green circles indicate DIG dominated fibers. Right: Histogram of the differences  $\Delta$ ${\rm E(B-V)}$ $=$ $\widehat{\rm  E(B-V)}$$_{\rm rim}$ $-$  $\widehat{\rm  E(B-V)}$$_{\rm hole}$, see text for details.} 
\label{figure:DustMapHistHoles}
\end{center}
\end{figure*}

\begin{figure*}
\begin{center}
\resizebox{0.5\hsize}{!}{\includegraphics[angle=000]{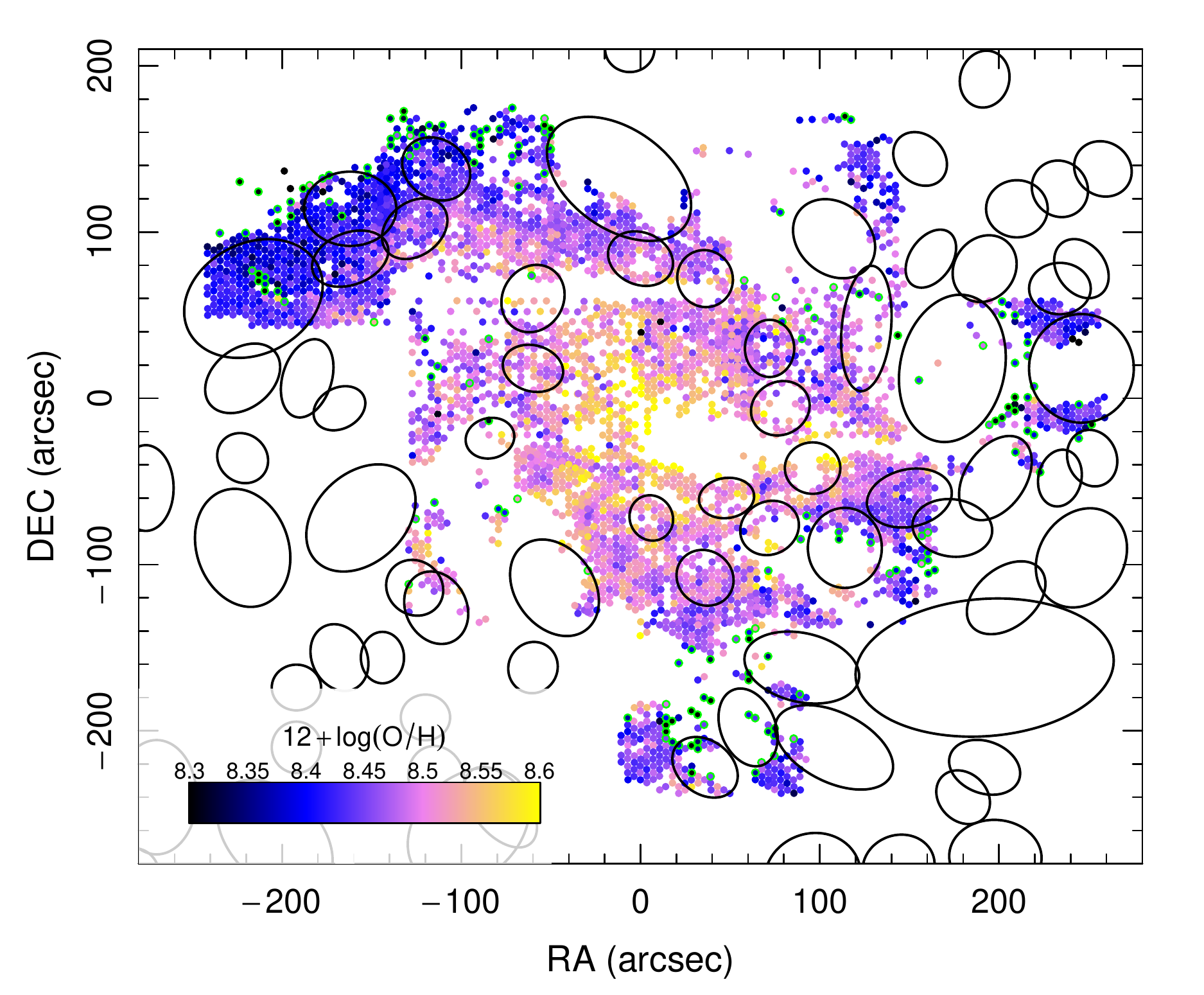}}
\resizebox{0.45\hsize}{!}{\includegraphics[angle=000]{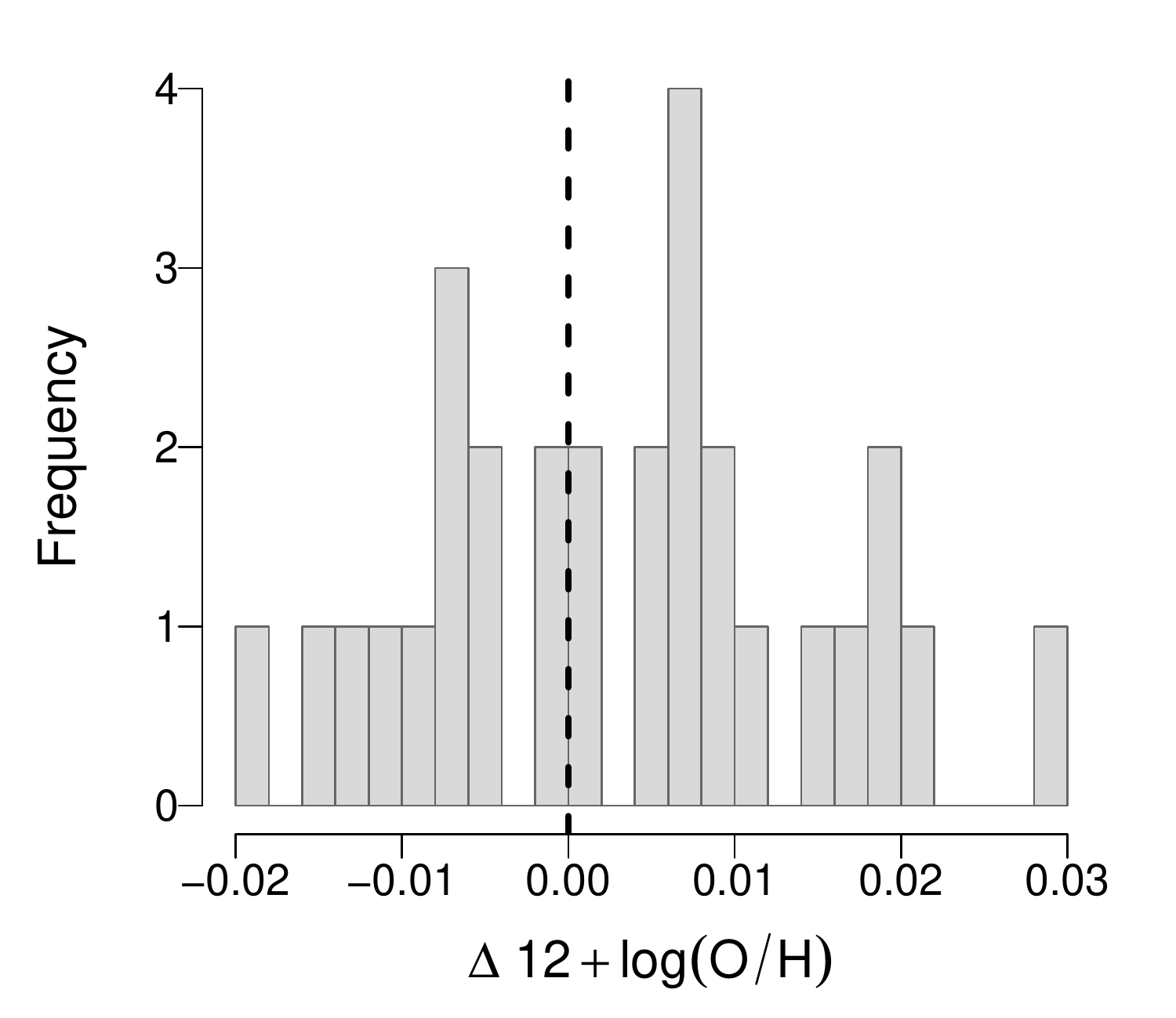}}
\caption{
  Left: H\,{\sc i} holes (black solid ellipses) superimposed on the 12+log(O/H) map of  NGC~6946. The small green circles indicate DIG dominated fibers. Right: Histogram of the differences  $\Delta$ ${\rm 12+log(O/H)}$ $=$ $\widehat{\rm  12+log(O/H)}$$_{\rm rim}$ $-$  $\widehat{\rm  12+log(O/H)}$$_{\rm hole}$, see text for details.} 
\label{figure:MetMapHistHoles}
\end{center}
\end{figure*}


\begin{figure*}
\begin{center}
\resizebox{0.5\hsize}{!}{\includegraphics[angle=000]{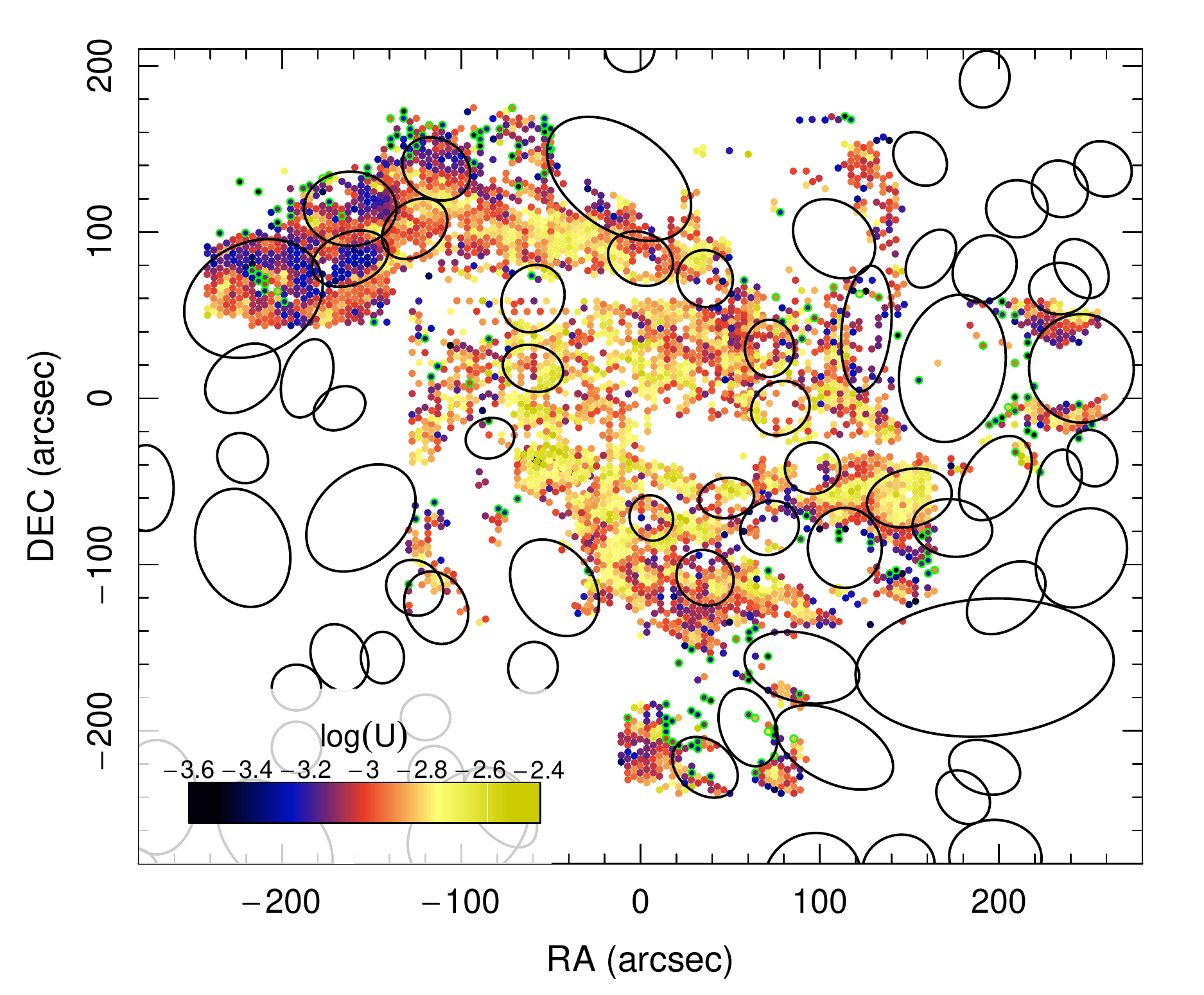}}
\resizebox{0.45\hsize}{!}{\includegraphics[angle=000]{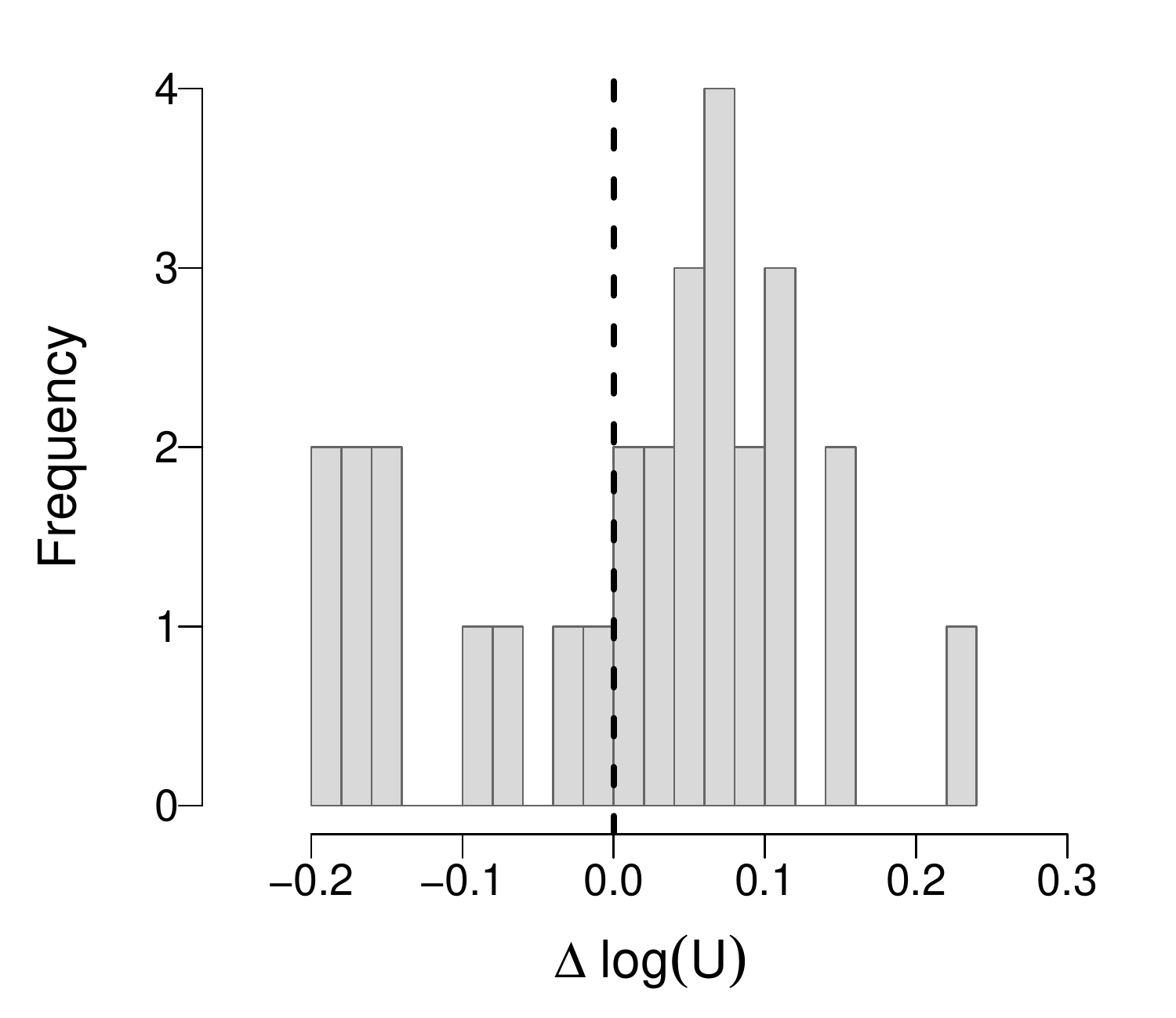}}
\caption{
  Left: H\,{\sc i} holes (black solid ellipses) superimposed on the log(U) map of  NGC~6946.  The small green circles indicate DIG dominated fibers. Right: Histogram of the differences  $\Delta$ ${\rm log(U)}$ $=$ $\widehat{\rm  log(U)}$$_{\rm rim}$ $-$  $\widehat{\rm  log(U)}$$_{\rm hole}$, see text for details.} 
\label{figure:IonMapHistHoles}
\end{center}
\end{figure*}

As indicated in the introduction, \citet{Boomsma2007} and \citet{Boomsma2008} identified and characterised 121  H\,{\sc i} holes, most of which
are located in the inner regions where the gas density and the star formation rate are higher. 
The origin of these H\,{\sc i} holes is analyzed in detail in \citet{Boomsma2008}. In general, they found that smaller holes  up to 1 kpc in size, are more likely to be caused by SN feedback. However, it is still a puzzle that many  H\,{\sc i} holes in NGC 6946 are observed without progenitor remnants. Also, they observed high velocity H\,{\sc i} gas, that is likely produced by the blown-out (into the halo) of the superbubbles that produce the holes in the disk, in agreement with the galactic fountain model \citep{Collins02}. In addition, larger holes could be a blend of a number of smaller ones, or even be caused by the combination of thermal and gravitational instability \citep{Dib05}.
Since the likely origin for most of these H\,{\sc i} holes is related to superbubbles created by multiple supernovae explosions, in this section, we explore such connection.

In Fig.~\ref{figure:hole-snr} we show the distribution of  H\,{\sc i} holes and recent supernovae remnants across the image of NGC~6946. The sizes and orientations of the H\,{\sc i} holes are indicated by the black solid ellipses, while the green points show the positions of the SNRs.  For this study, we use the sample of SNR candidates from \citet{Long2019}, who identified 147 candidates, using narrowband images, based on their elevated  [S\,{\sc ii}]$\lambda\lambda$6716+6731/H$\alpha$ ratios \citep{Mathewson1973}.

Inspection of Fig.~\ref{figure:hole-snr} suggests that  recent SNRs tend to avoid the H\,{\sc i} holes, in fact, SNRs are mostly distributed in the rims of the H\,{\sc i} holes. 
To quantify this effect, we estimated two distances: (a) the distance from the SNR to the closest H\,{\sc i} hole center (R$_{\rm SNR}$), and (b) the distance from the  H\,{\sc i}  hole center to the rim of the hole in the direction of the SNR (R$_{\rm rim}$). A histogram of the differences between distances (a) and (b) is shown in  Fig.~\ref{figure:histogram-distholesSNR}.  Every SNR is counted only once, and with respect to the closest hole center.  The peak around zero in this histogram indicates an excess of SNRs coincident with the rims of the H\,{\sc i} holes. The skewness of the histogram towards positive differences highlights that only few SNRs lie within the H\,{\sc i}  holes.
 Since the inclination angle of NGC 6946 is $\sim$32.6$\degr$ \citep{Bonnarel1988,deBlok2008}, projection effects could in principle alter the distribution of the histogram in Fig. \ref{figure:histogram-distholesSNR}. However, we estimate a maximum $\sim$15 \% correction in the obtained distances.

 On the other hand, the H\,{\sc i} holes are not related to the BPT classification of the individual fibers. Instead, different BPT classifications take place within and outside the H\,{\sc i} holes, as shown in Fig. \ref{figure:bptmap}. Also, the same figure shows that the DIG does not tend to dominate regions inside the holes, with only a $\sim$10$\%$ of DIG residing within the holes.


\subsection{The physical properties of H\,{\sc i} holes and rims}\label{SecHolesMaps}


 The excess of SNRs associated with the H\,{\sc i} holes rims is evidence that the induced star formation takes place in the shells of some superbubbles created by  multiple supernova explosions at their centres dozens of Myr ago. 
Figure \ref{figure:SFRMapHistHoles} (left) shows the log($\sum$$_{\rm SFR}$)  map of NGC 6946 together with the H\,{\sc i} holes superimposed.
A close examination of this map suggests that the  $\sum$$_{\rm SFR}$ tend to increase  towards the H\,{\sc i} hole rims. 
To examine this effect we estimated two  $\sum$$_{\rm SFR}$ for the whole sample, ({\it i}) the median star formation surface density inside each hole   $\sum$$\widehat{_{\rm SFR_{hole}}}$, which corresponds to the median values of the fibers from the center of the hole to up to 79\% of the  distance from the center to the edge of the ellipse that define each hole, and ({\it ii}) the median  for the holes rims  $\sum$$\widehat{_{\rm SFR_{rim}}}$, which corresponds to the median value of the fibers within 80\% to 130\%  of the distance from the center to the edge of the ellipse of each hole. The same selection of fibers inside the holes and in the rims is followed in the next sections, although the number of holes analyzed will  slightly change due to the different number of fibers used. The use of different distances to define center and rim fibers do not change our results drastically.

The difference $\Delta$ log($\sum$$_{\rm SFR}$) $=$ log($\sum$$\widehat{_{\rm SFR_{rim}}}$) $-$  log($\sum$$\widehat{_{\rm SFR_{hole}}}$)  is estimated only when more than 8 fibers are available in both, inside and in the rims of each hole, giving us a total of 31 holes. 
The histogram of these differences is shown in Figure \ref{figure:SFRMapHistHoles} (right), where it is clear that the histogram is skewed towards larger differences. From our sample, $\sim$67.7$\%$ of the holes show  an enhanced $\sum$$_{\rm SFR}$ in their rims.  To further quantify the degree of difference between rims and holes, we estimated a paired t-test, and obtained a t = 2.85, and a p-value = 0.007, meaning that there is a  0.7$\%$ probability that our result  happened by chance, hence we can reject the null hypothesis. 
When we discard the fibers previously identified as DIG (small circles in Fig. \ref{figure:SFRMapHistHoles}) and perform the same exercise, our sample is reduced to 28 holes, while the SFR enhancement in the rims holds, with $\sim$64 $\%$ of the difference showing a $\Delta$ log($\sum$$_{\rm SFR}$) $>$ 0. 


Following the same methodology, we analyzed the color excess E(B-V) as an indicator of dust extinction. We use the same sample used for the SFRs. The obtained  E(B-V) map is shown in Fig. \ref{figure:DustMapHistHoles} (left).  We analyze the color excess inside the holes and in the rims for 31 holes. The histogram of the difference $\Delta$ E(B-V) $=$  $\widehat{\rm E(B-V)}$$_{\rm rim}$ $-$ $\widehat{\rm E(B-V)}$$_{\rm hole}$, is shown in Fig. \ref{figure:DustMapHistHoles} (right). The color excess histogram indicates a skewness towards higher values in the rims, with a total of $\sim$61$\%$ of holes showing a higher color excess in their rims.
We run a paired t-test, and obtained a t = 2.17, and a p-value = 0.03, hence, our data suggest higher dust extinction on the hole's rims.
If fibers identified as DIG are not considered, the percentage of  $\Delta$ E(B-V)  $>0$ remains at $\sim$61$\%$.

We did the same exercise for the oxygen abundance and ionization parameter. Since the oxygen abundance and ionization parameter have the most restrictive criteria, we were able to analyze only 29 holes. The map of 12+log(O/H) is shown in Fig. \ref{figure:MetMapHistHoles} (left), while its respective histogram of the difference $\Delta$ 12+log(O/H) $=$  $\widehat{\rm 12+log(O/H)}$$_{\rm rim}$ $-$ $\widehat{\rm12+log(O/H)}$$_{\rm hole}$ in Fig.  \ref{figure:MetMapHistHoles} (right). The histogram indicates a minor shift towards higher abundances, with $\sim$58.6$\%$ of our data showing higher differences. The paired t-test indicate a  t = 1.45, and a p-value = 0.16. Hence, the difference in oxygen abundance between holes and rims is minor. Our results do not change when we exclude DIG dominated fibers (green circles in Fig. \ref {figure:MetMapHistHoles}). 


Similarly, the ionization parameter is displayed in Fig. \ref{figure:IonMapHistHoles} (left), while its respective histogram of the difference  $\Delta$ log(U) $=$  $\widehat{\rm log(U)}$$_{\rm rim}$ $-$ $\widehat{\rm log(U)}$$_{\rm hole}$, is shown in Fig. \ref{figure:IonMapHistHoles} (right).
The t-test indicate that we cannot reject the null hypothesis with t = 0.72, p-value = 0.48. Hence, we conclude that there is not a statistically significant difference in the ionization parameter between holes and rims. However, the histogram is skewed towards higher ionization parameter differences, with $\sim$66$\%$ of the holes showing positive differences.


From our statistical analysis, both the t-test and the skewness in the histogram show that the SFR has the strongest difference between the holes and the rims. To a lesser degree, E(B-V) shows as well a clear enhancement on the rims.
 On the other hand, the t-tests do not show a statistical significant difference between the chemical abundance and the ionization. However, both show a skewness in their histogram towards positive differences, specially higher for the ionization, with a $\sim$66$\%$ of the holes showing an  enhancement on the rims.
 


\section{Discussion}

 In this section we discuss the physical implications of the overabundance of SNRs in the rims of the H\,{\sc i} holes. As indicated in Section \ref{SecHolesMaps}, a remarkable feature in NGC 6946 is the existence and classification of 121  H\,{\sc i} holes \citep{Boomsma2008}. Our data indicate an overabundance of SNRs located around the rims of the H\,{\sc i} holes (Fig. \ref{figure:histogram-distholesSNR}). The  H\,{\sc i} holes in NGC 6946 were analyzed in detail by  \citet{Boomsma2008}, and they concluded that the creation of some of the holes can be attributed to the expansion of superbubbles generated by multiple SN explosions \citep[e.g., ][]{Dib06,Bagetakos11,Dib21}. In addition, they found that stellar feedback in the form of a galactic fountain is probably the origin of some of the H\,{\sc i}  holes. This is in agreement with the lack of bright emission at multiple wavelengths inside the holes, suggesting that the H\,{\sc i} holes are already devoid of gas, which could also explain the lack of SNRs located inside the holes.

As shown in Section \ref{SecHolesMaps}, we observe a clear enhancement in SFR and dust extinction on the hole's rims, and to a minor degree, the gas metallicity and ionization parameter. The star formation  is a complex process that is affected by internal and external mechanisms, as well as the aforementioned properties to different degrees.
For instance, since the existence of dust in galaxies affects the star-formation activity, it is not surprising that the observed enhancement in SFR is accompanied by a higher dust extinction in the hole's rims.  Indeed, it has been shown that dust grains increase the molecular formation rate by two orders of magnitude compared to the case without dust   \citep[][]{Hollenbach79}, drastically enhancing the star formation activity \citep[e.g., ][]{Asano13}. Hence, it makes sense that  bursts of star formation are associated with areas of high dust extinction.  

 Furthermore, the excess dust we observe in the hole's rims is likely produced by the SNRs. It is known that SNe  are very efficient dust producers (mainly core-collapse Type II-P, i.e., there is a plateau in their light curve), as well as  dust destroyers  \citep[i.e., the pressure of the ISM gas that is shocked by the expanding SNe blast wave generates a reverse shock that propagate through the expanding ejecta, partially destroying the newly formed dust, ][]{Bianchi07,Marassi19}. 
The composition and size of  grains formed in the SN ejecta that survive the subsequent passage of the reverse shock is still an open question. Recent models suggest that only between 1 to 8$\%$ of the observed dust mass produced by SNe will survive and contribute to the ISM \citep[e.g., ][]{Bocchio16, Sarangi18}. Taking into account the cumulative effect of  the high number of SNR in NGC 6946, it is not surprising we observe a higher dust extinction in the rims of the holes.

On the other hand, the ionization parameter in the rims of the holes shows only a small difference with respect to the holes, although  $\sim$66$\%$ of the rim's show higher values.
The ionization parameter directly reflects the ratio of radiation to gas pressure, or the interaction between the ionizing source and ionizing gas. 
Therefore, the ionization parameter increases with the luminosity of the stellar population and the hardness of the ionizing radiation field. 

It has also been suggested that high ionization parameters are the result of high SFRs, which lead to a larger reservoir of ionizing photons  \citep[e.g., ][]{Hainline09, Kewley13}. This would be in agreement with the observed enhancement in SFR in the rims of the holes. Hence, the observed difference is likely a result of the enhanced SFR taking place.
Another possible effect at play is the dust absorption. On galactic scales, \citet{Yeh12}  suggest that the ionization parameter can be suppressed by selective dust absorption of ionizing photons in the regions where U is highest, decreasing the observed ionization parameter. This suggests that the  ionization parameter we observe could be higher. 


Finally, we observe only a minor increase in the oxygen abundance in the rims of the holes. Considering that the chemical enrichment produced by the SNe will be mixed into an already enriched ISM (12+log(O/H) $\sim$ 8.5, see Fig. \ref{figure:snrohsv}), the cumulative effect of the SNe on the ISM metallicity is expected to be small and within the errors.

\section{Conclusions}

In this paper, we analyze the spiral galaxy  NGC~6946, also known as the Fireworks galaxy due to its extremely high number of SNRs. We link the high number of SNRs with  H\,{\sc i} holes in NGC~6946, and analyze their optical properties such as the oxygen abundance, SFR surface density, ionization parameter, and the BPT classification. 
 This work relies on the combination of the SNR  candidates of \citet{Long2019}, the H\,{\sc i}  holes catalog of \citet{Boomsma2008}, and the GMS IFU spectroscopy of Metal-THINGS for NGC 6946.  Our findings can be summarized as follows.

-The oxygen abundance of fibers hosting SNRs are in agreement with the rest of our data and follow the general tendency of the metallicity gradient of fibers not hosting SNRs. Altogether, this suggests that SNRs do not   contribute significantly to the radiation of the fibers.

  - Our data indicate an overabundance of SNRs located around the rims of the H\,{\sc i} holes (Fig. \ref{figure:histogram-distholesSNR}).  This is in agreement with the lack of bright emission at multiple wavelengths inside the holes, suggesting that the H\,{\sc i} holes are already devoid of gas, which also could explain the lack of SNRs located inside the holes. 

 - The overabundance of SNRs in the H\,{\sc i} hole rims indicates an enhanced star formation in the very recent past in those regions. Indeed, our IFU data shows a higher star formation rate on the rims compared to  within the H\,{\sc i} holes (see Fig. \ref{figure:SFRMapHistHoles}). This is  connected with the overabundance of SNRs, in the sense that expanding shells would compress the surrounding gas and trigger the  formation of a new generation of stars.  
Supporting this effect, our data indicates that the dust extinction is also higher in the rims. Indeed, this result is expected since dust grains increase the molecular formation and enhance the SFR.


- Our analysis provides a hint of an enhanced ionization parameter in the rims for 66$\%$ of the analyzed holes, although this is not statistically significant according to the paired t-test (p-value of 0.48). Additionally, our analysis shows only a marginal chemical enrichment, with $\sim$58.6$\%$ of holes showing a higher oxygen abundance in the rims.


Altogether, our data supports a picture in which multiple SN explosions occurred dozens of Myrs ago, producing the H\,{\sc i} holes, and compressing the gas and inducing star formation in the rims of the H\,{\sc i}  holes. 

\begin{acknowledgements}
We thank Dr T.A. Oosterloo  for providing us with the  H\,{\sc i} holes catalog.
Our team acknowledges funding from the International Space Science Institute (ISSI) for collaboration meetings.
L.S.P acknowledges support in frame of the program ``Support for the development of priority fields of scientific research''
of the NAS of Ukraine (``ActivPhys'', 2022-2023).   
M.A.L.L. acknowledge support from the Spanish grant PID2021-123417OB-I00.
M. R. acknowledges support from the grants CONACYT CF-86367; CY-253085, and DGAPA-PAPIIT (UNAM) IN 109919.
O.L.C acknowledges travel support from INAOE and Sistema Nacional de Investigadores (SNI).
This research has made use of the NASA/IPAC Extragalactic Database, which is funded by
the National Aeronautics and Space Administration and operated by the California Institute of Technology. 
We acknowledge the usage of the HyperLeda database (http://leda.univ-lyon1.fr).

\end{acknowledgements}

%
%

\end{document}